\documentclass[]{emulateapj}
\usepackage{graphicx}
\usepackage{natbib}
\DeclareGraphicsExtensions{.jpg,.pdf,.png,.eps,.ps}

\def\Michigan{1}
\def\Munich{2}
\def\ExcellenceCluster{3}
\def\MPE{4}
\def\UNDakota{5}
\def\LANL{6}

\shorttitle{A Parametrized Galaxy Catalog Simulator for Testing Cluster Finding}
\shortauthors{Song et~al.}

\begin{document}
\title{A Parametrized Galaxy Catalog Simulator For Testing Cluster \\
 Finding, Mass Estimation and Photometric Redshift Estimation \\
 in Optical and Near Infrared Surveys}

\author{
Jeeseon Song\altaffilmark{\Michigan}, 
Joseph J. Mohr\altaffilmark{\Munich,\ExcellenceCluster,\MPE}, 
Wayne A. Barkhouse\altaffilmark{\UNDakota}, 
Michael S. Warren\altaffilmark{\LANL},
and Cody Rude\altaffilmark{\UNDakota}}

\altaffiltext{\Michigan}{Department of Physics, University of Michigan, 450 Church St. Ann Arbor, MI 48109, USA}
\altaffiltext{\Munich}{Department of Physics, Ludwig-Maximilians-Universit\"{a}t, Scheinerstr.\ 1, 81679 M\"{u}nchen, Germany}
\altaffiltext{\ExcellenceCluster}{Excellence Cluster Universe, Boltzmannstr.\ 2, 85748 Garching, Germany}
\altaffiltext{\MPE}{Max-Planck-Institut f\"{u}r extraterrestrische Physik, Giessenbachstr.\ 85748 Garching, Germany}
\altaffiltext{\UNDakota}{Department of Physics \& Astrophysics, University of North Dakota, Grand Forks, ND 58202, USA}
\altaffiltext{\LANL}{Theoretical Division, Los Alamos National Laboratory, Los Alamos, NM 87545, USA}

\begin{abstract}

We present a galaxy catalog simulator which turns N-body simulations with Friends-of-Friends halo catalogs and IsoDen subhalos into mock, multiband photometric catalogs.  The simulator assigns galaxy properties to each subhalo in a way that reproduces the observed cluster galaxy halo occupation distribution, the radial and mass dependent variation in fractions of blue galaxies, the luminosity functions in the cluster and the field, and the color-magnitude relation in clusters.  Moreover, the evolution of these parameters is tuned to match existing observational constraints.  Field galaxies are sampled from existing multiband photometric surveys of similar depth using derived galaxy photometric redshifts.  Parametrizing an ensemble of cluster galaxy properties enables us to create mock catalogs with variations in those properties, which in turn allows us to quantify the sensitivity of cluster finding to current observational uncertainties in these properties.

We present an application of the catalog simulator to characterize the selection function of a galaxy cluster finder that utilizes the cluster red-sequence together with galaxy clustering on the sky.  Completeness varies with mass and redshift, and contamination roughly constant at around 35\% out to $z\sim0.6$.  We estimate systematic uncertainties due to the observational uncertainties on our simulator parameters in determining the selection function using five different sets of modified catalogs.  Our estimates indicate that these uncertainties are at the $\le15$\% level with current observational constraints on cluster galaxy populations and their evolution.  In addition, we examine the utility of the $B_{gc}$ parameter as an optical mass indicator and measure the intrinsic scatter of the $B_{gc}$--mass relation to be approximately log normal with $\sigma_{\log_{10}M}\sim0.25$.  Finally, we present tests of a red sequence overdensity redshift estimator using both simulated and real data, showing that it delivers redshifts for massive clusters with $\sim$2\% accuracy out to redshifts $z\sim0.5$ with SDSS-like datasets.
\end{abstract}


\section{Introduction}
\label{introduction}

Connecting baryonic cosmic structures with dark matter in the universe is very important in understanding the evolution of large scale structure.  This is the key to understanding the implications of large observational surveys within the context of theoretical and numerical studies of large-scale structure evolution.  Within this field of study clusters of galaxies have long been recognized as important laboratories for, e.g.,  studies of galaxy evolution \citep{faber07,hansen09}, intracluster medium enrichment \citep[e.g.,][more references therein]{lin04b} and critical sign posts for studies of the evolution of the large scale structure and cosmology through properties such as their mass function \citep[e.g.][]{white93,vikhlinin09,vanderlinde10} and their clustering \citep[e.g.][]{miller01,huetsi10}.  

Galaxy cluster surveys are becoming one of the key tools for unveiling the nature of  the dark energy \citep{wang98,haiman01}.  For example, \citet{gladders07} recently used a moderate scale galaxy cluster survey to constrain cosmological parameters using the self-calibration method \citep{hu03,majumdar03, majumdar04}.  Other analyses using Sloan Digital Sky Survey \citep[SDSS][]{york00} have used a large ensemble of nearby clusters to study cosmology \citep{rozo10}.  Although these optical surveys delivered large numbers of clusters, the cosmological constraints were only moderately strong.  This points towards the need to better understand the systematics of cluster surveys.  The key areas of concern include the cluster selection and the cluster mass estimation.  Without continued progress in these areas we will not be able to capitalize on the rich cosmological information emerging from, for example, the South Pole Telescope Sunyaev-Zel'dovich effect survey \citep{staniszewski08}, the Dark Energy Survey and the eROSITA X-ray survey \citep{predehl10}.

Many simulations have successfully reproduced observational features in clusters and galaxies, such as the luminosity function or the correlation function.  This usually has been done in one of two ways: using the framework of halo occupation distribution (HOD) \citep[e.g.,][]{berlind02, kravtsov04}, which can also depend on luminosity \citep[e.g.,][]{yang03} and by using a semi-analytical model for galaxy evolution in merger history trees \citep[e.g.,][]{kauffmann99}.  In the HOD framework, galaxies are populated within halos assuming a certain HOD and described by a conditional probability P\((N \mid M)\) that a halo with mass M contains N galaxies \citep{berlind02}.  \citet{yang03} extended the HOD modeling by labeling the galaxies with luminosities that use the conditional luminosity function (CLF) \(P(N \mid M)dL\), which gives the probability of finding a galaxy with luminosity in the range of L $\pm$ dL/2 as a function of halo mass M.  These methods are powerful tools to populate dark matter particle simulations with galaxies, but one must rely on the adopted shape of the HOD or a specific galaxy evolution model.  Certain aspects of galaxy evolution models, in particular, are still quite uncertain.  
 
A more recent scheme to create mock catalogs uses the Hubble Volume Simulation \citep{evrard02}\footnote{http://www.physics.lsa.umich.edu/hubble-volume}, where dark matter particles were chosen to be galaxies based on their local density, regardless of the host halo mass, and luminosities were assigned by the observed luminosity-density relation in SDSS \citep{wechsler04}.  
Another recent example of relating luminosities and galaxies was developed by \citet{vale06} using a subhalo catalog.  In that study, they introduced a non-parametric model to relate the luminosity of galaxies and the mass of the dark matter halo or subhalo which hosts it, under the assumption that the luminosity-mass relation is monotonic.  Their scheme, however, did not reproduce the observed LF in \citet{lin04a}.

Our objective is to provide simulated galaxy catalogs that reproduce many of the properties of the large scale structures in the universe, so that we can use these catalogs to analyze and calibrate various automated optical and NIR galaxy cluster analysis tools.  Moreover, we want our catalog simulator to allow us to turn the current observational uncertainties into uncertainties on the selection function and our ability to estimate masses in the optical.  To this end we assign the subhalo catalogs from the N-body simulation to be galaxies, and then directly implement observational features of galaxies in various wavebands without assuming any HOD, semi-analytical model, or luminosity-mass relation. 

This paper is organized as follows.  In \S \ref{nbody sim}, we describe the N-body simulation, including the creation of our subhalo catalog.  In \S \ref{mock catalog}, we introduce our method for assigning galaxy properties to subhalos.  In \S \ref{VTP}, a cluster finder is tested against the mock catalogs and analyzed.  In \S \ref{Bgc}, we discuss the reliability of the optical mass indicator, the $B_{gc}$ parameter.  Finally in \S \ref{discussion}, advantages and limitations of our approach are discussed, as well as future directions of the project.  Throughout the paper, we assume the cosmological parameters to be $\Omega_{M} = 0.3,  \Omega_{\Lambda}=0.7$, and the Hubble parameter to be $H_{0}=70$ km $s^{-1}$ Mpc$^{-1}$.

\section{N-body Simulation}
\label{nbody sim}

The catalog simulator uses subhalos from an N-body simulation to determine the locations and dynamics of galaxies.  Below we describe the simulation and extraction of the halo and subhalo catalogs.

\subsection{Halo Catalogs}

We carried out a cold dark matter (DM) only structure formation simulation with the Hashed Oct-Tree (HOT) algorithm, initially described in \cite{WarSal93a}, running at Los Alamos National Laboratory.  We modeled a universe based on the concordance cosmology \(( \Omega_{M} = 0.3,  \Omega_{\Lambda}=0.7, H_{0}=70 \) km \(s^{-1}\) Mpc\(^{-1}\)).  Initial conditions were derived from transfer functions calculated by CMBFAST \citep{seljak96}.  This DM simulation was performed with the same code and parameters as the series of simulations described in \cite{WarAbaHol06}.  The simulation followed a cubic computational volume $816.3 h^{-1}\rm Mpc$ across with $1280^3$ (2.1 billion) particles.  Each particle had a mass of \(3.63\times10^{9}\) M$_{\odot}$.

DM halos are identified by the Friends-of-Friends (FoF) method \citep{davis85}.  The FoF method identifies a set of spatially associated particles, roughly contained within an isodensity surface with a value of 
$$\rho(h_{link})=\frac{\alpha M_{p}}{(4\pi/3)h_{link}^{3}} ,$$
where \(M_{p}\) is the particle mass and  \(\alpha\) is a constant of order 2.  For each object, one can identify how many linked particles it contains.  Here the adopted linking length, $h_{link}$, is 0.2 times the average inter-particle spacing.

\subsection{Subhalo Catalogs}

Subhalos are identified independently from halos.  The subhalo finding algorithm, called the IsoDen method \cite{PfiSalSte97}, selects all density enhancements.  Subhalos are traced to the smallest clumps which contain at least 10 particles, corresponding to a  subhalo mass of $3.6\times10^{10} M_{\odot}$.  The IsoDen method calculates the spatial density field of particles to identify local peaks as centers of subhalos.  At each center, iso-density surfaces grow to find all particles that belong to the center until any two different surfaces start to touch each other.  

Subhalos are then paired up with host halos if they lie within the  virial radius r$_{200}$ of the host halo, within which the mean density is 200 times higher than the mean density of the universe.  Since the simulation output box is periodic, subhalos on the edges of the box are also checked for their membership with hosts on the other side of the box.  When a subhalo matches with two different host halos, we choose the more massive halo as the host.  Although FoF masses (that are given for the host halos) are not exactly same as spherical overdensity (SO) masses, \citet{lukic09} found that the agreement between the two masses is good (within 5\%) for most halos ($\sim$85\%) in their N-body simulations.  The other 20\% of the halos had clear substructures and non-relaxed halos.  We assume that for all of our halos, the agreement between FoF mass and SO mass is good enough for our purpose.  We note that, however, this is one of the sources of scatter in the following comparison.  

\begin{figure}[ht]
\begin{center}
\includegraphics[scale=0.45]{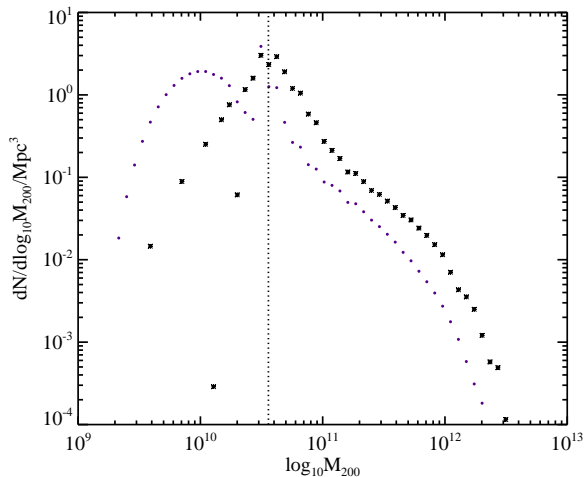}
\end{center}
\caption{Subhalo mass function:  We plot the number density of subhalos per $\log_{10}M_{200}$ as a function of subhalo mass for the field subhalos (dots), scaled by $\frac{200}{\Omega_M}$, and the  cluster subhalos (stars).  We define all halos with masses $M_{200}>5\times10^{13} M_\odot$ as clusters.}
\label{subhalo properties}
\end{figure}

We examine the matched subhalo catalog to see whether the basic properties such as the subhalo mass functions (SMF) and the radial distributions are consistent with observations of cluster galaxies.  For example, to compare the mass function of the field population with that of cluster members, we scale the field SMF by $\frac{200}{\Omega_{M}}$, because the density in cluster DM is by definition enhanced by 200 times compared to the field within the virial region. 
Here we take clusters to be those halos with mass greater than $5\times 10^{13} M_{\odot}$ that contain member subhalos that were matched with the halos within $r_{200}$ (i.e., no galaxy--free clusters are allowed).  The field population includes lower mass halos along with their member subhalos (mass less than the above mass cut), as well as subhalos that were not matched to any halo.  
 In Figure~\ref{subhalo properties}, filled circles represent the field population and stars represent cluster members.  For this plot, the subhalos were not cut off by their definition in mass (i.e., 10 particles or more or mass greater than $3.6\times 10^{10}M_{\odot}$.

One characteristic to note is that there is a cutoff on the low mass end for subhalos in the field and in clusters that is driven by the mass resolution of the simulation, as denoted by the dotted line.  Note that the field SMF extends to roughly three times lower mass than the cluster SMF.  For the purposes of catalog simulation, we choose the mass threshold of subhalos at $3\times10^{10}M_{\odot}$, which allows us to use the bulk of the available population of cluster subhalos.  
The difference between field and cluster SMF turnover is an indication that subhalos must have been destroyed in dense regions, and particle groups of lower mass than this threshold are not found in sufficient numbers as subhalos.  

Also in Figure~\ref{subhalo properties}, we find an overdensity of cluster subhalos as compared to field subhalos after scaling by a factor of $\frac{200}{\Omega_m}$.  In the $K_s$-band, the normalization offset between the cluster luminosity function \citep{lin04a} and the field luminosity function \citep{kochanek01} is about $\sim$3.5, which is consistent with the offset at a mass range around $5\times10^{11}M_{\odot}$ in the subhalo mass function if we assume differences in $K_s$ band mass to light between the field and the cluster are not significant.  
Note that \citet{gao04} find an offset in the subhalo abundance distribution in massive halos as compared to that in the Universe as a whole.  Our result is also consistent with \citet{giocoli08}, who report that the normalization of the \emph{unevolved} subhalo mass function is higher than the SMF in the host halos by a factor of two.

\begin{figure}[ht]
\includegraphics[scale=0.45]{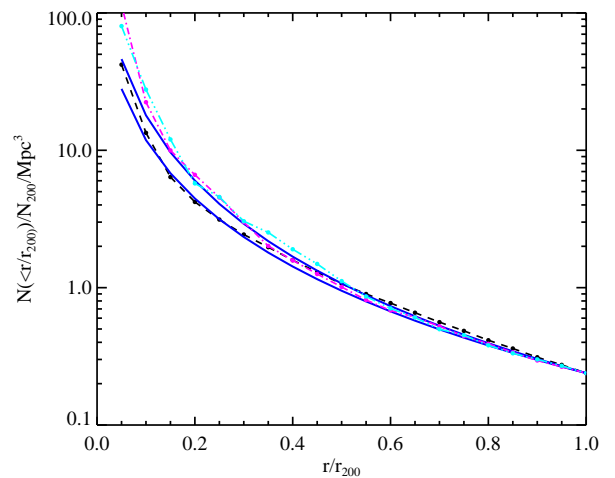}
\caption{Subhalo radial profiles:  We plot the density of subhalos as a function of distance $r/r_{200}$ from the center of their parent halo.  The profiles have been normalized by the total number of subhalos within each host halo's virial radius $r_{200}$.  Halos in the mass range of [5e+13,2e+14], [2e+14,8e+14],[1e+15,3.5e+15] in units of $M_{\odot}$ are shown with cyan dash-dot-dot, magenta dash dot, and black dash lines, respectively.  For comparison we plot an NFW model with a concentration $c=5$ (the upper blue solid line) and $c=3$ (the lower blue solid line) for the DM particle distribution.}
\label{subhalo profile}
\end{figure}

In Figure~\ref{subhalo profile} we show subhalo radial profiles compared with an NFW profile \citep{navarro97}.  A total of 12,000 halos are stacked to produce this figure and Figure~\ref{subhalo properties}. 
For the radial profile in Figure~\ref{subhalo profile} these halos are then separated into three mass bins, where only first 3,000 halos are shown.  The lowest mass bin, shown with the dashed line, represents halos with mass between $5\times 10^{13} M_{\odot}$ and $2.0\times 10^{14} M_{\odot}$, and the next mass bin, shown with the dash-dot line, contains halos with mass between $2.0\times 10^{14}$ and $10^{15} M_{\odot}$ and the highest mass bin contains all remaining halos that reach a mass up to $3.5\times10^{15} M_\odot$.  We compare these radial profiles of the subhalos with the NFW profile (solid lines), with a concentration of 5 and 3.  These three subhalo profiles are in reasonable agreement with the NFW profile, but the halos in the highest mass bin, represented with the dash line in the figure, show a deficit of subhalos in the central region when it's compared to the NFW profile with concentration of 5 (for DM particles).  This deficit has already been noted in previous studies \citep[e.g.,][]{deLucia04, ghigna00} and is referred to as `antibiased' relative to the dark matter in the central regions of the halos.  \citet{deLucia04} suggested that this is naturally explained as a consequence of the orbital decay combined with dynamical friction and mass loss that subhalos experience in high density regions.  Subhalos that are on orbits that take them through the cluster center are quickly destroyed and soon are no longer distinguishable from the central halo.  Also, observed galaxy radial profiles in clusters are generally well fit by an NFW model,  but with a lower concentration of 3 or 4 \citep[e.g.][]{lin03b,lin04a}.  In our catalog generator these DM subhalos are used as galaxies, and so any radial properties of the subhalos will be reflected in the resulting galaxy population.  For testing cluster finders that do not assume a specific shape for the cluster galaxy profile, this mismatch between the radial profiles of the cluster galaxies and the subhalos is not crucial.  

\section{Catalog Simulator}
\label{mock catalog}

The catalog simulator works on the halo and subhalo catalogs described in the previous section.  
A 2-D sky view from an observation is a projection of a lightcone where the volume depends on redshift extent, cosmological parameters and the angular size or field-of-view.  Ideally, one would like to build a mock catalog from a large enough simulation to include self-consistently the cosmic expansion and structure formation evolution along the lightcone.  In the absence of such a simulated light cone, one can use one or more outputs from a periodic simulation box.  In the tests described in the sections below we use a single $z=0$ simulation output and stack it as needed to simulate the more distant universe.  Therefore, the simulations used in this paper do not naturally follow the evolution of the underlying dark matter large scale structure.

To build our lightcone we set an observer at a random position inside the box with a random line-of-sight direction -- additional boxes are located along the chosen line-of-sight to simulate the higher redshift universe.  This random realization makes it possible to build different mock catalogs from the same simulation.  Cluster and galaxy redshifts are assigned using their comoving distances from the observer and their peculiar velocities.  The angular diameter distance together with the field of view determines the portion of the simulation that is included in the simulated catalog at each redshift.  Once redshifts are determined, galaxy properties are assigned to reproduce several observed quantities, including the luminosity function, the color distribution and the halo occupation number (HON).  In the following sections,  we describe how observational constraints are used to guide the catalog simulator.

\subsection{Galaxy Luminosity}

In principle we could use the subhalo mass to assign a luminosity.  However, the mapping from dark matter mass to stellar mass (and therefore luminosity) is complex, because of the stripping processes that affect the DM halos and stellar populations differentially and that depend on the orbits of the galaxies through the cluster.   Therefore, rather than do this, we assign galaxy luminosity randomly to each subhalo in a manner that reproduces the observed luminosity function in the appropriate environment.  We treat the BCG separately.  Because there is no strong mass segregation of subhalos, these two approaches should produce similar results.  Moreover, with our approach we parametrize the field and cluster galaxy populations using Schechter luminosity functions, and we can vary those populations systematically to explore the sensitivity of cluster finding and mass estimation to these properties.  In addition, with this approach we have the flexibility to change the number density of field galaxies relative to cluster galaxies by altering the luminosity parameters.  

When assigning luminosities to galaxies,  we follow the observed $K_s$-band LF as given by \citet{kochanek01} for the field population and \citet{lin04a} for cluster members (for parameter values see Table~\ref{primaryLF}).  Both studies found a Schechter function as a good fit with the faint end slope fixed for local samples.  Within the observed uncertainties, we match the two LFs by adopting $K_*$ with the corresponding faint end slope, $\alpha$ (fixed).  Luminosities of BCGs are determined using the observed relation between $L_{BCG}-M_{200}$\citep{lin04b};
\begin{equation}
\frac{L_{bcg}}{10^{11}h_{70}^{-2}L_{\odot}}=4.9\pm0.2\left(\frac{M_{200}}{10^{14}h_{70}^{-1}M_{\odot}}\right)^{0.26\pm0.04}
\end{equation}

\begin{table}[bp]
	\begin{center}
	\caption{2MASS LF parameters}
	\vspace{0.05in}
		\begin{tabular}{l || c | c }\hline\hline
		& $\textbf{\(M_{K\ast}\)}$ & $\textbf{\(\alpha\)}$ \\ \hline
		Cluster$^{(1)}$ & $-24.34 \pm 0.01$ & -1.1\\
		Field$^{(2)}$ & $-24.16 \pm 0.05$ & -1.1  \\
		\hline
	\end{tabular}
	\label{primaryLF}
	\end{center}
	\footnotesize{$^{(1)}$\citet{lin04a}; $^{(2)}$\citet{kochanek01}.  
	}
\end{table}

We impose a flux limit in drawing random luminosity from a LF that corresponds to the limit where the number of subhalos would match the expected number of galaxies in the luminosity function.  This flux limit is determined differently for the clusters and the field.  For the clusters, we compare the  HON in the simulated clusters (the number of subhalos within each host halo) to the HON in real clusters found by \citet{lin04a};
\begin{equation}
N_{200}=36\pm3\left(\frac{M_{200}}{10^{14}h_{70}^{-1}M_{\odot}}\right)^{0.87\pm0.04},
\end{equation}
where \(N_{200}\) is the number of galaxies within \(R_{200}\) with an absolute K-band magnitude of -21.0, approximately $M_*+2.5$.  
This comparison provides a corresponding flux-limit.  

\begin{figure}[ht]
\includegraphics[scale=0.45]{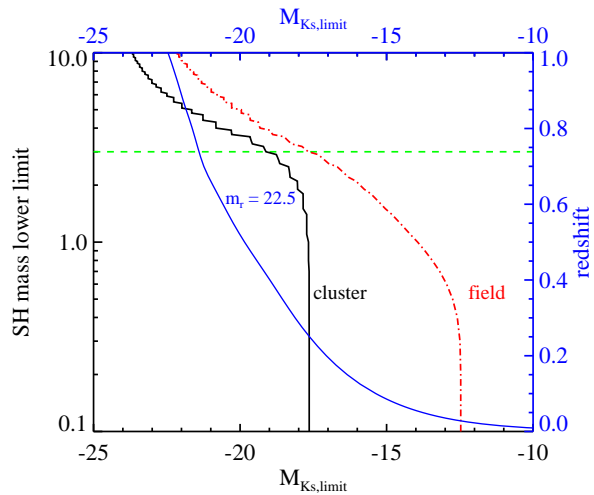}
\caption{A plot of how deep one can go down in simulating a survey, depending on underlying subhalo populations.  Black x-, y- axes with the black solid line (clusters) and red dot-dashed line (field) show how different lower mass threshold of subhalos to be included in a mock catalog determines $K_s$- limit by changing subhalo number density in the simulation.  The cluster and field lines are essentially the subhalo mass function curves in the absolute $K_s$ magnitude space.  Blue axes with blue solid line show a SDSS-like survey depth in $r$- band at 22.5 corresponds to which absolute magnitude in $K_s$- at different redshift.  With subhalo lower mass limit of $3.0\times10^{10}M_{\odot}$ indicated by the green dashed line, the survey limit in $K$-band determined by subhalo density in clusters is deep enough at redshift greater than $\sim$0.2, as the green line intercepts with cluster and field mass function curves. }
\label{Klim}
\end{figure}

For the field population, we find the survey depth from the total number density of subhalos above a certain mass cut (i.e., $3 \times 10^{10} M_{\odot}$) in the simulation box by comparing it to the number density of all galaxies (by integrating the LF to the survey depth) found by \citet{kochanek01}.  The resolution of this subhalo catalog is sufficient to push this survey depth down to -17 or -18, deep enough for a SDSS-like survey for all but the very lowest redshifts (Figure~\ref{Klim}), as the intercepting magnitude in absolute $K_s$-band ($\sim$-17.5) between the adopted subhalo mass cut ($3.0\times10^{10}M_{\odot}$ shown with the green line) and the field subhalo number density in absolute $K_s$ space is deeper than a SDSS-like survey limit in absolute $K_s$ at redshift of 0.2 and beyond.  The vertical flat region is due to the subhalo mass limit in the current simulation that we are using (i.e., there are no subhalos with mass less than $1.0\times10^{10}$M$_{\odot}$ in clusters). 

In determining the flux limit we evolve the $K_*$ of both field and cluster LFs with redshift according to the BC model (see \S\ref{galaxy colors}).  This is equivalent to an assumption that $K_*$ galaxies in $\it{both}$ clusters and the field are red galaxies (i.e., no recent star formation), which is reasonable because $K$-band light is less affected by recent episode of star formation.  We explore the effects of different field versus cluster population evolution in Section~\ref{modifications} below.

In the process described above we also use the observed HON redshift evolution.  \citet{lin06} examined the N-M relation evolution by combining their local sample with a sample of several dozen clusters extending to redshifts $z\sim1$.   Modeling the evolution as 
\begin{equation}
N(M,z)=N(M,z=0)(1+z)^{\gamma_{HOD}}.
\end{equation}
they showed that the data currently suggest very weak evolution of the HON.  We adopt this form for the HON evolution with a fiducial $\gamma_{HOD}=0$ and vary this parameter to explore the impact of variations in the HON evolution on cluster finding.

\subsection{Galaxy Type}
\label{galaxy type}

Recent studies \citep{weinmann06} have measured the fraction of blue galaxies within groups and clusters as a function of halo mass, halocentric radius, and luminosity using
a group and cluster catalog that was optically selected from SDSS DR 2 in the redshift range between 0.01 and 0.2 \citep{yang05}.  Results indicate that the blue fraction is higher in the outskirts of halos and in less massive halos.   \citet{gerke07} extended the measurement of the blue fraction to higher redshift and to the field.  They measured the cluster blue fraction as a function of radius from the cluster center, showing that the blue fraction gradually increased with  distance from the center, approaching the blue fraction in the field.

This provides a convenient and simple way to model the galaxy populations in our simulated catalogs.  We adopt an approach where clusters and the field contain only two types of galaxies -- blue and red, with blue being star-forming galaxies and red being passively evolving galaxies.  The variations in color of each of these types of galaxies are described in the next section.  We further assume that clusters contain a special type of red galaxy - a brightest cluster galaxy (BCG)-- which is the most luminous galaxy that is typically located near the center of the cluster. 

We construct a functional form for the blue fraction to include variation with halo mass $M_{200}$, distance from the halo center $\frac{r}{r_{200}}$, and galaxy absolute magnitude in $K_s$-band  $M_{K_s}$ by assuming variations in blue fraction associated with these three parameters are separable:
\begin{equation}
f_{blue}(z)=A(M_{200})f_{b}(\frac{r}{r_{200}})f_{b}(M_{K_s})(1+z)^{\gamma_{blue}},
\label{bluefrac equation}
\end{equation}
where $A(M_{200})$ is a parameter associated with halo mass, $f_{b}(\frac{r}{r_{200}})$ is the blue fraction as a function of $\frac{r}{r_{200}}$, $f_{b}(M_k)$ is the blue fraction as a function of its $K_s$ magnitude and the redshift evolution is assumed to be a power law with index $\gamma_{blue}$.  The adopted blue fraction as a function of luminosity is measured in $r$-band, but because the primary band in our scheme is $K_s$, we assume an arbitrary fixed color $r$-$K_s$ of 4.0 in using the published results to calibrate our model.  To find the parameter $A(M_{200})$, we integrate the blue fractions over the radial profile and the luminosity function within the virial region.
Once $A(M_{200})$ is determined for halo masses from $\log_{10}M_{200}=11.85$ to $\log_{10}M_{200}=15.55$, each galaxy is assigned a probability of being blue using Equation~\ref{bluefrac equation} together with the halo mass, the subhalo position within the halo and the assigned $K_s$ band luminosity.  Beyond the virial radius we have a field  population, which is taken to have a blue fraction of 80\%. 

Since \citet{bo84} presented evidence that the fraction of the blue galaxies in clusters increases with redshift (the BO effect), many studies have attempted to replicate their work through different star-formation indicators \citep[e.g.,][]{kodama01,poggianti06}.  There are, however, many complications involved in attempting to define the evolution of the fraction of blue galaxies \citep[see][and references therein]{gerke07}.  A recent study on the evolution of the blue fraction in groups and in the field \citep{gerke07}, measured the blue fraction in groups as a function of redshift between 0.75 and 1.3.  They found a nearly constant blue fraction in groups in this redshift range, and a rising field blue fraction beyond $z=1$.  Since there is no single self-consistent observational study on the evolution of blue fractions in clusters at different redshifts ranging from local to the very distant universe, we parametrize the blue fraction evolution using $\gamma_{blue}$ shown above, which allows us to explore the impact of more or less rapid evolution in the galaxy population.

\subsection{Galaxy Color}
\label{galaxy colors}

Once every galaxy is assigned a $K_s$-band magnitude and an identity as either a blue or red galaxy (note that BCGs are modeled as red galaxies), colors in optical and near-IR bands follow.  Red galaxies have a characteristic spectrum with no strong emission lines and a prominent 4000~$\AA$ break since their stars have been passively evolving since their last episode of star formation at high redshift.  We use a Bruzual \& Charlot model \citep[BC03;][]{bc03} to assign the colors for the red galaxy population.  We assume a single burst of star formation at $z=3$ and let the galaxies passively evolve to $z=0$ according to a single stellar population (SSP) synthesis model.  We use six different models with six distinct metallicities corresponding to different luminosities, allowing us to construct a tilted red-sequence.  These SSP models are then tuned in metallicity so that the tilt of the color-magnitude relation for the Coma cluster ($z=0.0234$) is reproduced.  The models provide apparent magnitudes in $ugrizJHK_s$ bands at specific redshifts out to $z= 2.98$ and at six different luminosities at each redshift.  We interpolate using these models to generate model colors at  intermediate redshifts and luminosities. 
Since each of the models contain only a single population of stars, we introduce scatter in the metallicity-luminosity relation to model variations among galaxies (i.e., to introduce intrinsic scatter in the color-magnitude relation).  The observationally motivated scatter is 0.075 mag \citet{barkhouse06}.

The fact that blue galaxies with current or recent star formation activity are have in general complex star formation histories (e.g., O'Connell et al. 1997) makes them more difficult to model.  Therefore, we sample colors of blue galaxies directly from a subset of the FLAMEX\footnote{FLAMINGOS Extragalactic Survey is a deep imaging survey covering 7.1 \(deg^{2}\) sky in the J and \(K_s\) filters.  The purpose of this survey is to study galaxy and galaxy cluster evolution at \(1<z<2\) \citep{elston06}} database.  This database contains photometry in the $BRIJK_s$ bands, as well as for the $\it{Spitzer}$ IRAC [3.6] [4.5] [5.8] [8] micron filters, and measured photometric redshifts of 175,000 galaxies.  We choose only those galaxies with well measured fluxes in NIR bands, leaving us with about 36,000 galaxies.  To couple a galaxy with representative colors, we divide the adopted subset of FLAMEX into further subsets according to redshift and $K_s$-band luminosity.  For each simulated galaxy, we choose the subset which is closest to its redshift and $K_s$-band luminosity, and then randomly select an observed galaxy from that group, with the additional constraint that it be bluer than the corresponding halo's color-magnitude relation at that redshift.

\subsection{Observational Effects}

Once magnitudes are determined for each galaxy, observational uncertainties are applied to this model universe to produce a realistic simulated catalog.  The galaxy catalog simulator can choose a specific telescope or instrument with a specific survey depth to add noise to the galaxy magnitudes.  An integration time is determined such that the corresponding signal-to-noise ratio, which includes both systematic and statistical uncertainties, reproduces the measured uncertainties for the selection of telescope or instrument.  The $S/N$ is given by,
\begin{equation}
\frac{S}{N}=\frac{\sigma_{src}^2}{\sqrt{\sigma_{src}^2 +\sigma_{bkg}^2 + \sigma_{sys}^2}},
\label{SN}
\end{equation}
where $\sigma_{src}$, $\sigma_{bkg}$ are uncertainties in measured flux of a source and sky brightness (as a background), respectively, while $\sigma_{sys}$ is the systematic uncertainty, which is band-dependent for a specific instrument.  The chosen survey depth will impose a magnitude cut on galaxies and typically our final catalogs contain galaxies with $S/N>5$ (i.e., $\sigma_{mag}\le0.2$).

\subsection{Variations in Galaxy Populations}
\label{modifications} 

One of the advantages an empirical catalog simulator like ours is that any input parameter, such as the adopted LF for clusters and the field or the evolution in the HON and the galaxy color distribution, can be changed to create a different mock catalog within the same simulation output.  Changing one parameter at a time to produce another mock catalog can be done, and comparing the results from the primary run to those from other runs with modifications allows us to test how much the parameter changes affect the catalog and the followon cluster finding or mass estimation.  This provides a straightforward way to examine the systematic uncertainty of the selection function for the cluster finder due to the uncertainties in the creation of the mock catalogs.  The general trend of the modifications is physically motivated, but the exact values for the deviations are not observed quantities.    In our model, there are five parameters that can be modified: the blue fractions as a function of physical properties of host clusters and galaxies,  the evolution of the fraction of red galaxies within clusters with redshift, the luminosity function of the field population, the evolution of HON with redshift, and the tightness of the color-magnitude relation.


\begin{figure}[ht]
\includegraphics[scale=0.45]{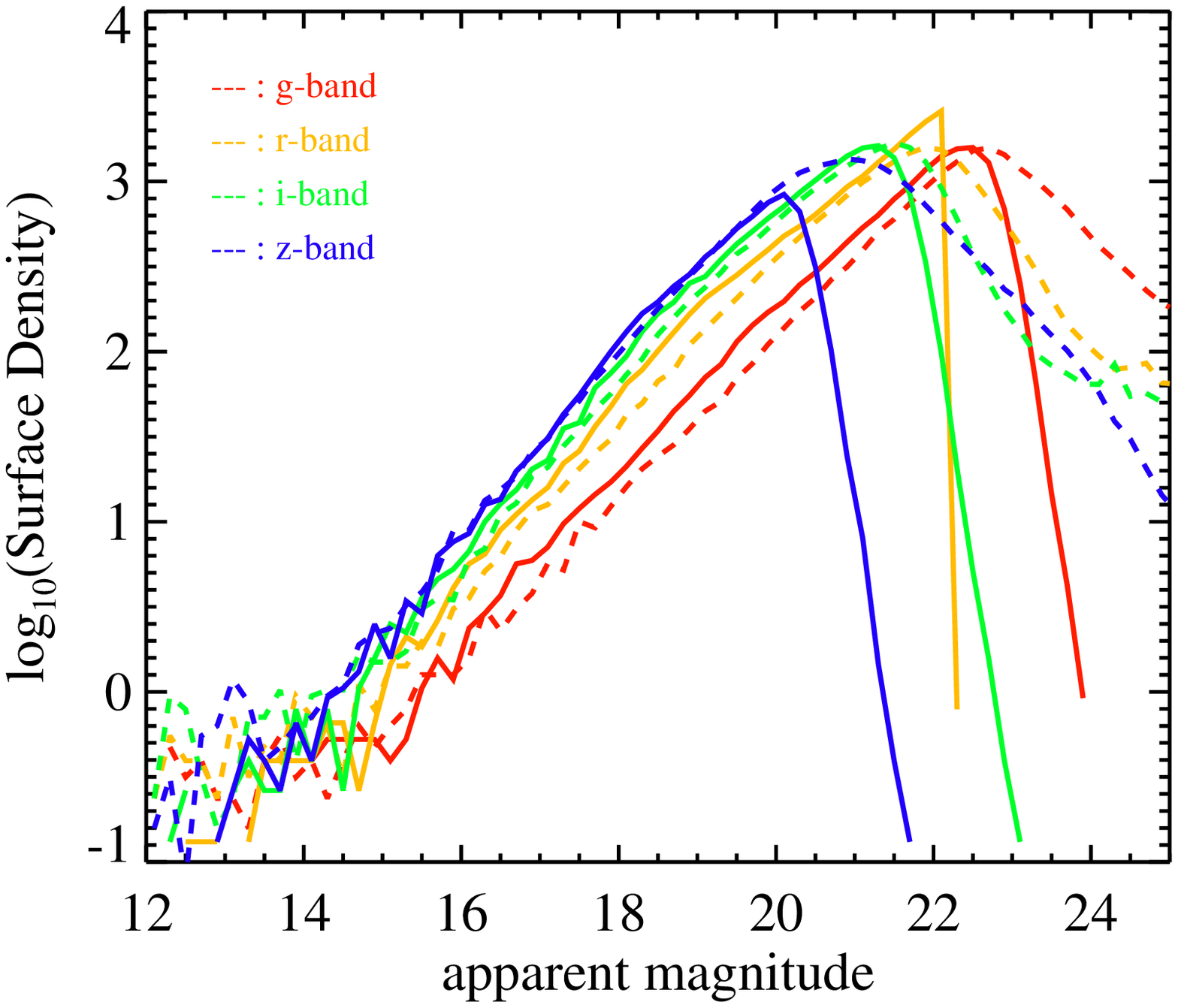}
\includegraphics[scale=0.45]{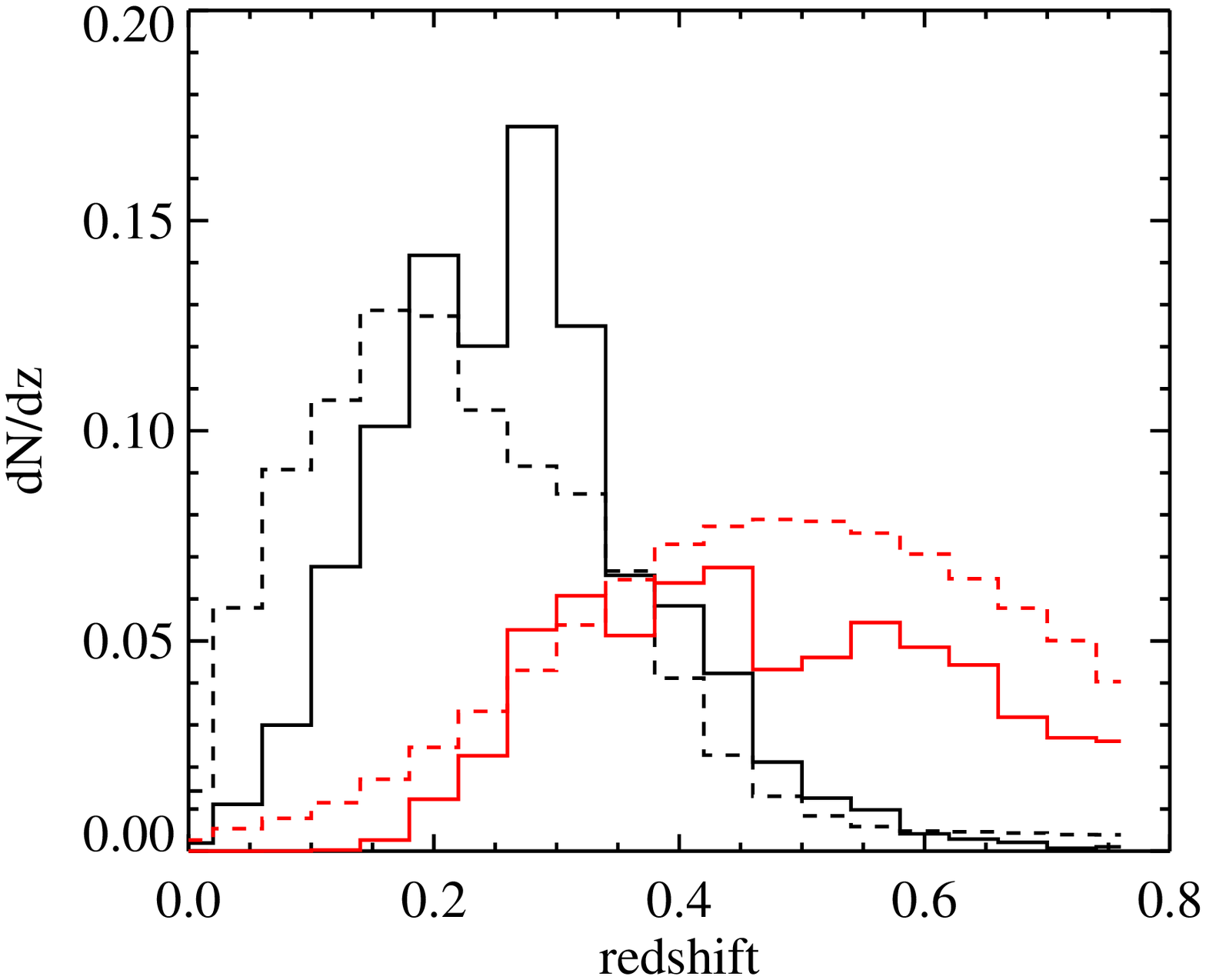}
\caption{Comparison in galaxy counts and redshift distribution:  \emph{Top}: Solid lines represent galaxy counts for the mock, with blue for $u$-, green for $g$-, yellow for $r$-, red for $i$- and black for $z$- bands. Dashed lines exhibit the same distribution for SDSS. \emph{Bottom}: The redshift distribution of the mock (solid lines) and  the measured photometric redshift distribution in SDSS DR7 (dashed line).  The black lines are galaxies brighter than 20 mag in $r$-band and the red lines are galaxies fainter than 20 mag in $r$-band.}
\label{logN-logS}
\end{figure}

\section{An SDSS--like Simulated Catalog}
\label{product}

We use the simulator to produce an SDSS--like galaxy catalog.  In the final galaxy catalog, we include positions, redshifts, peculiar velocities, photometric redshifts, magnitudes in $\it{ugriz}$ in the AB system, $JHK_s$ magnitudes and the $[3.6]$ and $[4.5]$ micron $\it{Spitzer}$ IRAC bands in the Vega magnitude system.  Table~\ref{primary parameters} shows the input parameter values for the primary run.  

\begin{table}[bp]
	\begin{center}
	\caption{SDSS-Like Catalog Parameters}
	\vspace{0.05in}
		\begin{tabular}{l|c|r}\hline\hline
		Property                          & Parameter                      & Value\\ \hline
		Field Luminosity Function                           & $K_{*}$                           & -24.21\\
		Cluster Luminosity Function                       &$K_{*}$                             & -24.33 \\
		Blue Fraction Evolution      & ${\gamma_{blue}}$      & 0.00 \\
		HON Evolution              &${\gamma_{HOD}}$     & 0.00\\
		Red Sequence scatter & $\Delta_{RS}$            & 0.07\\
		\hline
	\end{tabular}
	\label{primary parameters}
	\end{center}
\end{table}

The resulting galaxy catalog is then subjected to a series of validation tests which include logN-logS surface density for each band, color-magnitude relations for clusters, galaxy luminosity functions of clusters, halo occupation numbers, radial profiles, and blue fraction as a function of radius and mass.  Some of the basic features of this catalog appear in Figure~\ref{logN-logS}.  The top figure shows the logN-logS for both galaxies in the simulated catalog (solid lines) and those for $\sim12$deg$^2$ of the sky from the SDSS archive (dash lines).  The bright end for the SDSS galaxies is noisy due to the small sky area.  It is clear that the two distributions are in good agreement at magnitudes brighter than the flux limit of the simulation (at r$\sim22$).  The bottom figure shows the redshift distributions for the catalog compared to SDSS.  The solid lines show the distribution of assigned photometric redshifts in the simulation, while the two dashed line shows the measured photometric redshift distribution that is available in public SDSS DR7 data archive \citep{Abazajian09}.  Both solid lines and dashed lines are divided into two sets according to galaxies' magnitude.  The black lines toward the lower redshift is for brighter galaxies with $r$ magnitude cut at 20 or brighter, while the red lines toward the higher redshift is for galaxies fainter than $r$ of 20.  The general trend between the solid lines and the dashed lines agree with each other, but the SDSS galaxies show a somewhat smoother distribution.  We note that the depth of the two datasets are not identical, although this particular version of the mock catalog is generated to simulate an SDSS-like survey.  In the mock catalog, galaxies are cut off at the magnitude of 22.5 in $r$-band, while the SDSS galaxies used in this analysis are not.  Fainter galaxies must have larger photometric uncertainties which would make photometric redshift measurement less reliable. 

\begin{figure}[ht]
\includegraphics[scale=0.45]{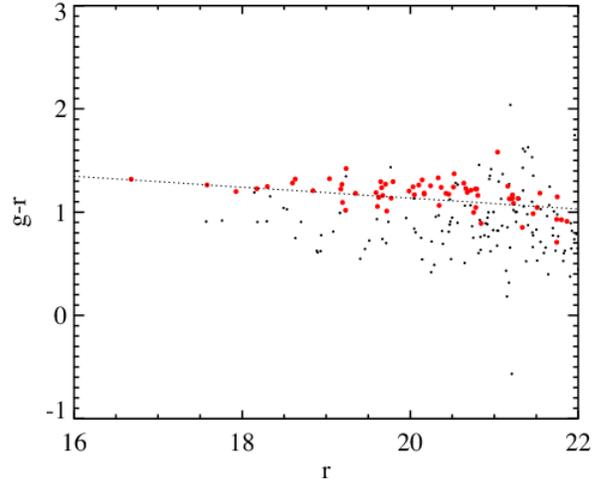}
\caption{Example of color-magnitude diagram.  This contains a cluster with mass of $4.5\times 10^{14} M_{\odot}$ at redshift 0.21.  Red dots are the cluster members.  The model red-sequence line at z=0.2 is shown.}
\label{cmr}
\end{figure}

Figure~\ref{cmr} shows an example of the color magnitude distribution of galaxies toward a cluster, which has mass of $4.5\times 10^{14}M_{\odot}$ and a redshift of 0.21.  The dotted line represents the red-sequence at redshift of 0.2 by the same galaxy models used in the galaxy simulator.   In order to test how well a simulated red-sequence represents a real cluster's red-sequence at a given redshift, we construct a redshift estimator based on the red-sequence of clusters, which in principle is comparable to the redshift estimator of the RCS group (see \citet{gladders05} for details), and measure redshifts of clusters based on the simulated catalog.  We use the same red galaxy models by BC03 that we use in the galaxy simulator described in \S\ref{galaxy colors}.  The detail of the tests of the redshift estimator is described in \S\ref{photo-z}.  In the tests, we measure comparable scatter in redshift estimates for simulated clusters as compared to our ensemble of clusters based on SDSS-$\it{Chandra}$ joint dataset below $z=0.6$.  This demonstrates that the simulated galaxy colors are consistent with the real galaxy colors in clusters with $z<0.6$.


\section{Evaluating a Red Sequence Cluster Finder}
\label{VTP}

In this section, we test a cluster finder - the Voronoi Tessellation and Percolation (VTP) cluster finding algorithm \citep{barkhouse06} using the simulated catalogs described above. This cluster finder code is still undergoing  development and testing to optimize the cluster detection parameters (Barkhouse et al. 2011, in preparation). In the VTP algorithm, clusters are detected as spatial overdensities by Voronoi Tessellation and Percolation method \citep{ramella01} within redshift shells defined using the expected cluster galaxy colors in the color-magnitude relation.  Because the finder searches in color-magnitude space, its output includes the redshift estimations as well as positions and richness measurements.  The VTP finder has been run on the SDSS DR6 to build a cluster galaxy catalog where these test results are utilized.  Completeness and false-positive (contamination) tests show how effectively this finder detects clusters in the simulation.  `Halos' refer to the dark matter halos in the simulation, while `clusters' refer to the systems that are recovered by the cluster finder.

The VTP algorithm first filters the galaxy catalog to select galaxies within 3$\sigma$ of the corresponding red-sequence at each redshift. The redshift-dependent red-sequence location in color-magnitude space is determined by assuming a stellar synthesis model with a star formation epoch with a burst at $z=5$ prior to passive evolution \citep{kodama97}.  Redshifts from the finder, therefore, can be biased by the assumed model.  In order to correct this bias, the raw prediction on redshift by the VTP finder is calibrated with redshifts in the mock catalog, since the mock catalogs use an SSP model with a different star burst formation time ($z=3$).

\subsection{Completeness $s(M_{200},z)$}

We analyze the VTP cluster selection results to measure the completeness and contamination of the resulting cluster catalog.  Note that completeness and contamination measures are sensitive to some degree to how the matching of the cluster and halo catalogs is done.  In this study, we match the VTP cluster candidates and the true dark matter halos by drawing a boundary for each dark matter halo with its $r_{200}$ and then selecting all VTP clusters that lie within this boundary.  In cases of multiple cluster--halo matches, we select the more massive halo as the true match.  The completeness $s(M_{200},z)$ is defined as the fraction of detected halos of that mass and redshift out of all the halos in the mock catalog with that mass and redshift.

\begin{figure}[ht]
\includegraphics[scale=0.45]{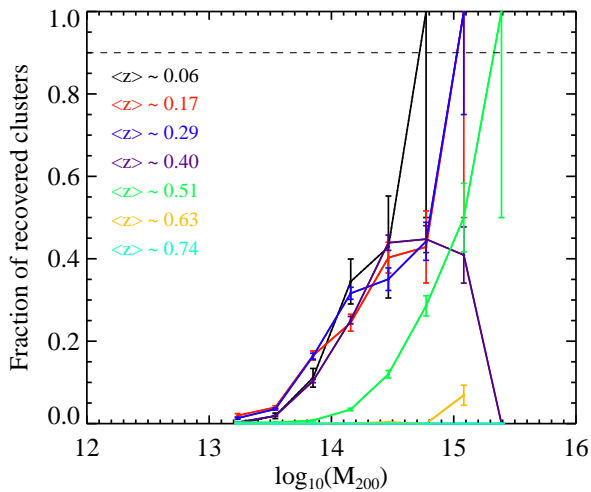}
\caption{A plot of the completeness $s(M_{200},z)$ versus cluster mass and redshift for halos that are recovered by the VTP finder.  Different redshift ranges are encoded using different colors.}
\label{completeness}
\end{figure}

Figure~\ref{completeness} shows the completeness of the primary run.  Note that the completeness is a strong function of mass with essentially a cutoff at some threshold mass that increases with redshift. The completeness in the lowest redshift bins ($<z>\sim$ 0.06) shows poorer performance than that in the next two bins ($<z>\sim$ 0.17 and 0.29), and this may be because these nearest clusters of galaxies are projected onto a much larger portion of the sky as compared to high redshift systems, effectively decreasing the signal to noise of a given system.  Note also that at these redshifts the limited number of subhalos does not allow us to populate the cluster and field populations to the full depth of the flux limited surveys, and this may also impact our results.  The error bars shown include only Poisson noise and therefore reflect only the statistical uncertainties on our completeness measurements.  

\subsection{Contamination $c(z)$}

\begin{figure}[ht]
\includegraphics[scale=0.45]{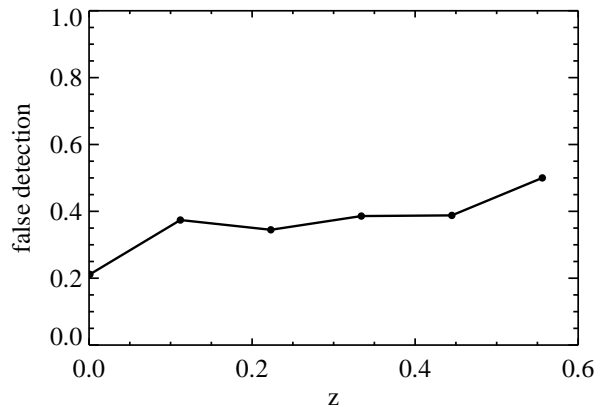}
\caption{We plot the contamination fraction $c(z)$ of all VTP clusters found that do not match to any true halo in the simulations.  At SDSS depths the VTP cluster catalog exhibits low contamination at $z<0.6$}
\label{contamination}
\end{figure}

We define the contamination $c(z)$ to be the fraction of detected VTP clusters not matched with dark matter halos.  To measure this we cross-match the VTP clusters with dark matter halos, requiring a separation of equal to or less than each dark matter halo's $r_{200}$.  Clusters with the smallest deviations in both spatial and redshift space are flagged as  the best candidates.  Clusters with no overlapping halos are taken as contamination.  The number of these false clusters relative to the total number of detected clusters determines the contamination.  Figure~\ref{contamination} shows the contamination fraction versus the redshift assigned to the contaminating cluster.  Clusters at all redshift where a SDSS--like survey probes, i.e., $z<0.4$, show a very steady contamination level at around the 40\% level.  Note that we do not examine the mass distribution of the contaminating clusters, because our only mass estimate in the case of contaminating clusters comes from the measured optical mass indicator $B_{gc}$, which is a poor indicator of cluster mass (see \S\ref{Bgc}).

\subsection{Estimating Systematic Uncertainties on Completeness and Contamination}

The completeness and contamination measurements presented above determine the best estimate of the selection function for the VTP finder, which can then be used in the cosmological interpretation of a VTP cluster catalog.  An important element of any analysis is to include not only the best estimate of the selection function but also the associated uncertainties.  These uncertainties derive from current observational uncertainties in physical properties of the cluster and field galaxy populations, and they are therefore systematic in nature-- e.g., an enhanced blue fraction at high redshift will make it systematically more difficult to select clusters at those redshifts.  We can use those observational uncertainties to characterize the uncertainties in the VTP selection function, and of course as the observational constraints on optical properties of the cluster and the field galaxy populations tighten we can improve our simulated catalogs and reduce the systematic uncertainties on the selection function of the cluster finder.  We proceed by creating several more simulated catalogs with input parameters varied around the best measured values.  Then we run the VTP finder on each of these modified catalogs to determine the selection in each case.  By comparing the selection from a modified catalog to that derived from the primary catalog one can estimate the effect of changing each simulation parameter on the performance of the finder.  Below we illustrate how this can be approached by carrying out such an analysis on the VTP cluster finder. 

\begin{figure}[htb]
\includegraphics[scale=0.48]{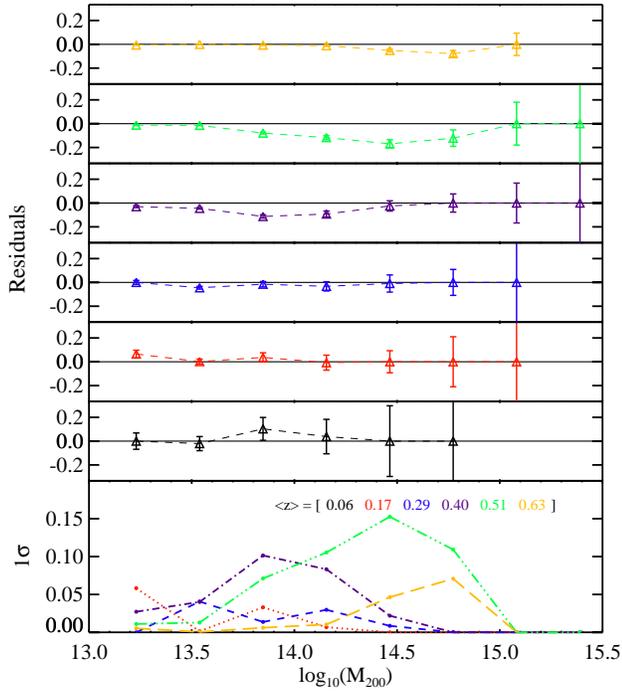}
\caption{The change in completeness due to changes in the blue fraction, which depends on the $K$-band luminosity, the mass of the host halo, and the galaxy's distance from the host halo center, as well as redshift evolution factor in the form of (1+z)$^\gamma_{blue}$, where $\gamma_{blue}$ is -1.6.  Results show the change as a function of mass for each redshift slice considered between the fiducial blue fraction model and one that is perturbed by 1$\sigma$ in its parametrization.  The bottom panel shows the systematic uncertainty in the completeness due to the 1$\sigma$ observational uncertainty in this blue fraction.}
\label{bluefrac}
\end{figure}

\begin{figure}[ht]
\includegraphics[scale=0.48]{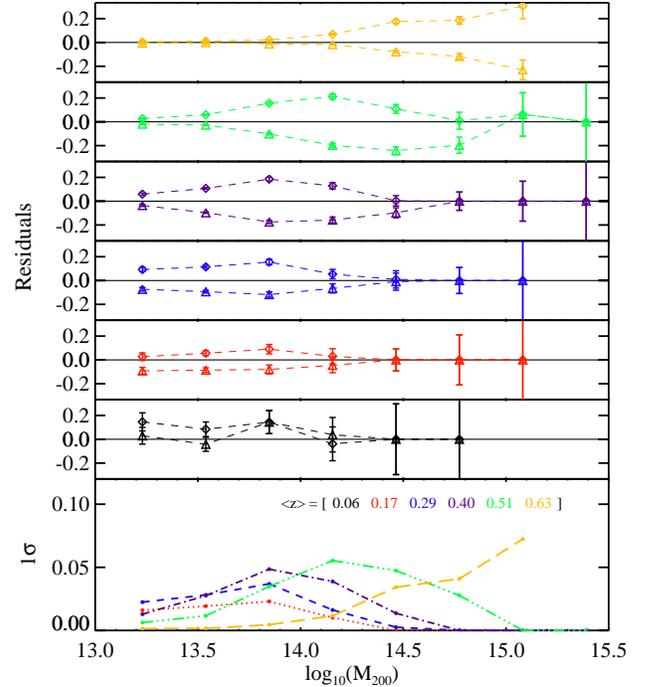}
\caption{The change in completeness due to changes in the evolution of the HON with  $\gamma$=+1 (diamond) or -1 (triangle).  Each panel with different color coded symbols represents different redshift bin of which the sequence is shown in the lowest panel.  The lowest panel shows the systematic uncertainty in the completeness due to the 1$\sigma$ observational uncertainty in HON.}
\label{hon}
\end{figure}

\begin{figure}[ht]
\includegraphics[scale=0.48]{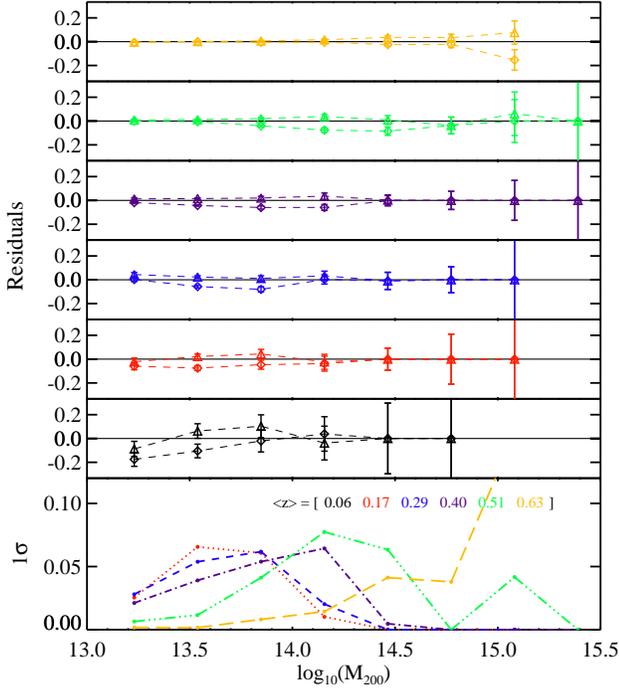}
\caption{The change in completeness due to changes in the intrinsic scatter of the cluster red-sequence  $\Delta_{RS}$=0.02-0.03 (blue) and $\Delta_{RS}$=0.12-0.15 (red).  Again, the bottom panel corresponds to the 1$\sigma$ uncertainty in the completeness due to the uncertainties in the intrinsic scatter.}
\label{rs scatter}
\end{figure}

\begin{figure}[ht]
\begin{center}
\includegraphics[scale=0.48]{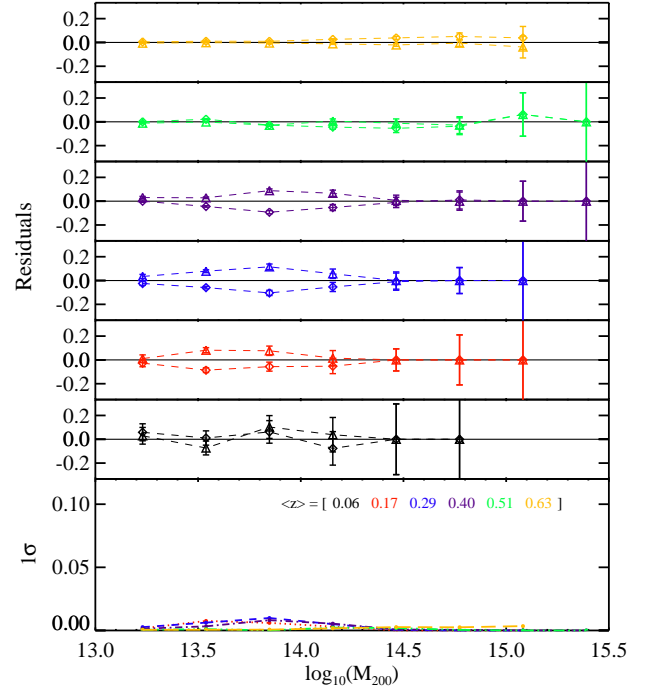}
\end{center}
\caption{The change in completeness due to changes in relationship between the field and cluster luminosity functions (LF).   The shows the case where the field LF is made fainter, and the red is the case where the field LF is made bright.  The bottom panel shows the 1$\sigma$ uncertainties in the completeness due to remaining uncertainties in the offset between the cluster and field LFs.}
\label{LF}
\end{figure}

The residual completeness, defined as the completeness of a modified catalog where the completeness of the primary run is subtracted, is shown in several figures: one for tests of the impact of blue fraction (Figure~\ref{bluefrac}), HON evolution (Figure~\ref{hon}), intrinsic scatter in the red sequence (Figure~\ref{rs scatter}), and relative luminosity of the field and cluster populations (Figure~\ref{LF}).  In each figure the bottom panel collects the results from each redshift, scales to the estimated 1$\sigma$ uncertainty on completeness due to this effect, and presents no uncertainties to improve the readability.

In Figure~\ref{bluefrac}, a modified catalog is generated with blue fraction as a function of $M_{200}$, $r_{200}$, and $M_{K_s}$ (see \S \ref{galaxy type}) and with redshift evolution included in the form of (1+z)$^\gamma_{blue}$.  The adopted $\gamma_{blue}$ is -1.6 so that the blue fraction at redshift of 1 to be 80\%.  Since the blue fraction depends on several factors, it's not possible to quote one value for the blue fraction at certain redshift.  The adopted $\gamma_{blue}$ is somewhat arbitrary in this presented exercise, since how blue fraction in clusters evolves with redshift is still controversial.  The fact that there are less red galaxies in clusters at high redshift, the selection becomes not as good as in the primary catalog test.  This is shown as dashed lines above the primary catalog completeness shown at zero. 
Figure~\ref{hon} shows the comparison in completeness for the HOD modification run where the evolutionary factor, $\gamma_{HOD}=\pm1$ is implemented.  With $+1$ shown with diamond, for example, there are more galaxies in a cluster of a given mass at higher redshift, and this makes it easier to detect these systems.  The triangles show the other case with $-1$, where there are fewer cluster galaxies at higher redshift, and so the finder detects fewer systems.  Figure~\ref{rs scatter} shows two different levels of intrinsic scatter in the red-sequence: one with smaller (blue) - $\sim$ 0.02 to 0.03 - and the other with larger scatter (red) - $\sim$0.12-0.15 - than the fiducial scatter in the primary catalog.  This test illustrates how the intrinsic scatter of the color-magnitude relation alters the performance of the cluster finder.  The smaller the scatter is, the more clusters are recovered by the finder, for example, since the finder relies on the existence of the red-sequence to isolate cluster galaxies from the field galaxies projected nearby.  In Figure~\ref{LF}, triangle (diamond) symbols represent the case where the $K_*$ of the field LF is brightened (dimmed) compared to that of the cluster LF.  The completeness shown in the lower redshift bins is counter-intuitive, where in high-z the change is too small to see the effect.  It is currently under investigation why the performance of a cluster finder gets better when the field galaxies get brighter, but a possible explanation is that brightened field galaxies at high z contribute to signals of clusters at lower z.  We also note that the changes are at a level of $<$1\% which could be random fluctuation.  Changing the field galaxies to make them brighter or fainter than the cluster does not have a significant impact on the completeness.   

\begin{figure}[ht]
\begin{center}
\includegraphics[scale=0.45]{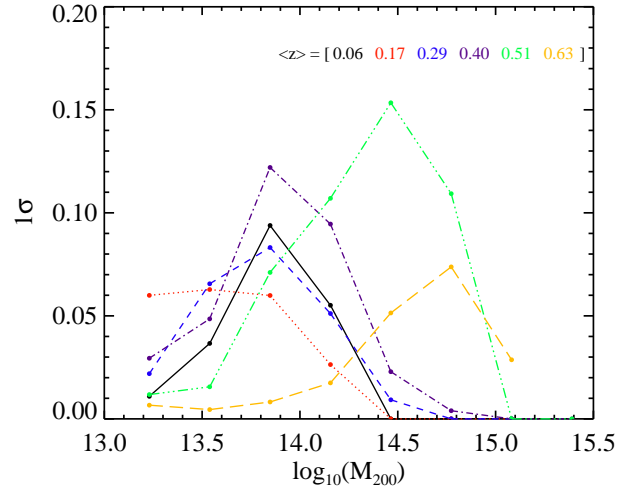}
\end{center}
\caption{Combined uncertainty in completeness as a function of mass and redshift.  Each 1$\sigma$ uncertainty from the different simulation parameter tests is combined in a quadrature sum assuming contributions from each parameter are independent.  Current observational uncertainties on the properties and evolution of cluster galaxy properties translate into typically $\sim10$\% systematic uncertainties in the survey selection function.}
\label{totalsigma}
\end{figure}

As noted above, the bottom panel of Figures~\ref{bluefrac}, \ref{hon}, \ref{rs scatter} and \ref{LF} shows the systematic uncertainty in the completeness due to observational uncertainties due to uncertainties in the blue fraction in clusters, the HON,  the size of the intrinsic scatter in the red sequence, and the field/cluster LF.  The runs used in Figures~\ref{bluefrac}, \ref{hon}, \ref{rs scatter} and \ref{LF} use characteristic values for the deviations from the best-fit values, which are large enough to see the effect of changes in the performance of the finder.  We can then use these measured changes in completeness to estimate the changes one would see for a $1\sigma$ shift in each of the simulation parameters.  Colors for redshift bins are the same as in Figure~\ref{completeness} and the others. Overall, it is clear that the existing observational uncertainties in the properties of cluster and field galaxy populations out to $z\sim0.8$ do not translate into large systematic uncertainties in the selection function for the VTP finder.  To fully characterize the uncertainties in the selection function would require an analysis of the covariances among these effects.  Here we are providing a demonstration of how one would proceed (Figure~\ref{totalsigma}).  In the limit of small covariances in the impact on the completeness of variations among these key simulation parameters, one can simply take the quadrature sum of the contributions to the uncertainty from each different parameter.  This leads directly to an estimate of the systematic uncertainty on the completeness as a function of mass and redshift that could be used in the cosmological analysis of any cluster sample extracted using the finder.  In a similar manner the impact of mock catalog simulator parameters on the contamination could be carried out.

\begin{figure}[ht]
\includegraphics[scale=0.45]{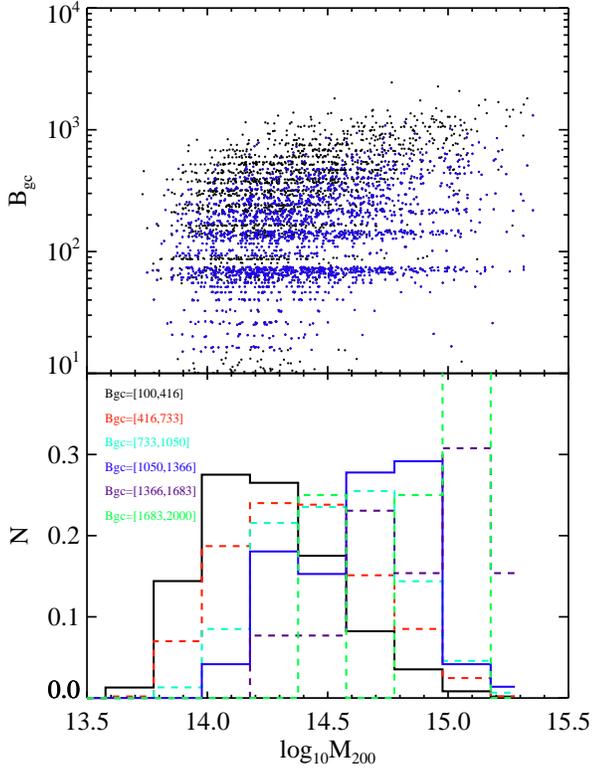}
\caption{The $B_{gc}$ versus halo mass relation.  \emph{Top}: red symbols are clusters at redshift greater than 0.4. \emph{Bottom}: The distribution in mass of clusters that lie within fixed ranges in $B_{gc}$.}
\label{Bgc scatter}
\end{figure}

\section{Performance of the $B_{gc}$ Mass Estimator}
\label{Bgc}

Along with finding clusters one must be able to reliably estimate their masses to be able to use the evolution of the mass function with redshift to study the nature of the dark energy or cosmic acceleration.  Establishing a reliable estimator in optical bands stands as a significant hurdle for optical cluster cosmology studies.  This may well be because, unlike in the X-ray and SZE where the signature of the hot gas within the virial region is dramatically enhanced by it's higher temperature and density, the galaxies change more gradually as they move from the field to the cluster.  In particular, red sequence galaxies exist not only in clusters, but also in groups and more generally as tracers of the large scale structure.  Because the red sequence galaxy population comes not only from the cluster virial region but also from the surrounding cluster environment, any signature derived from this population (i.e. the number of red galaxies or the integrated luminosity from this population) seems to be less well correlated with cluster virial mass than measured extracted from the X-ray and SZE.  Here we use our simulated catalogs to explore one well known optical mass estimator.

One measure of the strength or richness of the cluster galaxy population is using $B_{gc}$, which is the amplitude in the cluster center-galaxy correlation.  The $B_{gc}$ parameter, first pioneered by \citet{longair79} and later extensively tested by \citet{yee99}, is defined as follows;
\begin{eqnarray}
\label{Bgc equation1}
B_{gc}=N_{bkg}\frac{D^{\gamma-3}A_{gc}}{I_{\gamma}\Phi[M(m_0,z)]}, \\
\label{Bgc equation2}
A_{gc}=\frac{N_{net}}{N_{bkg}}[\frac{(3-\gamma)}{2}]^{\theta^{\gamma-1}}.
\end{eqnarray}
Here $N_{bkg}$ is the background galaxy count down to an apparent magnitude $m_0$ and $\Phi[M(m_0,z)]$ is the integrated LF of galaxies up to the absolute magnitude M corresponding to $m_0$ at the cluster redshift $z$.  $I_{\gamma}$ is a constant that depends on the choice of $\gamma$, and D is the angular diameter distance to $z$.  As shown in Equation~\ref{Bgc equation1}, $B_{gc}$ measurements depend on the LF that is adopted and the background number counts.  \citet{yee99} tested the sensitivity of $B_{gc}$ values to these two parameters and to the magnitude limit down to which the adopted LF was integrated and concluded that they did not strongly affect $B_{gc}$ values as long as the normalization of the LF was carefully measured.

In Figure~\ref{Bgc scatter} we plot the richness of the systems in the cluster catalog extracted using the VTP algorithm versus the real mass of the halos in the simulation.  As seen in our scatter plot of the $B_{gc}$ vs. $M_{200}$ for 8000 halos in the mock catalog (Figure~\ref{Bgc scatter}), the scatter in this relation is quite large and the apparent correlation is quite weak.  Red symbols in the plot, which represent clusters at redshift higher than 0.4, show higher redshift clusters tend to have lower $B_{gc}$ values for their mass.  
As shown in Equation \ref{Bgc equation2}, the measurement of $B_{gc}$ depends on the net count of galaxies through $A_{gc}$ ($N_{net}=N_{cluster \ members}-N_{bkg}$) which decreases with redshift due to the flux-limited nature of the mock catalog.  This results in the preference of a lower $B_{gc}$ for  higher redshift systems.  We also note that the $B_{gc}$ measurement by the VTP finder is restricted within the red-sequence slice for each cluster, which results in ``quantization'' in $B_{gc}$ values in the Figure~\ref{Bgc scatter} upper panel.  

\begin{figure}[ht]
\includegraphics[scale=0.45]{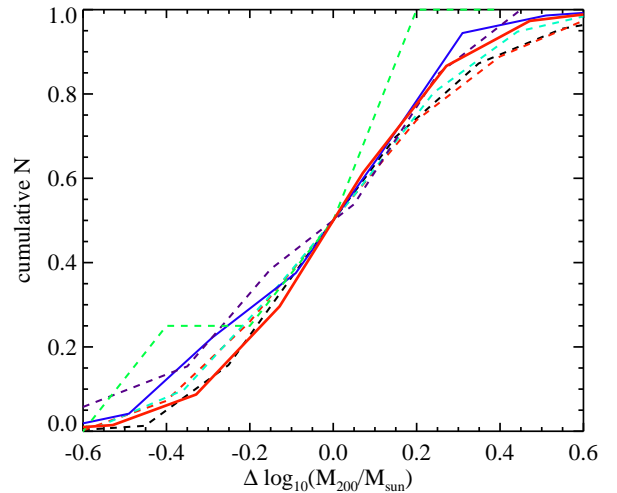}
\caption{The cumulative distribution in mass of different cuts in $B_{gc}$ shifted so that they overlap at the 50 percentile point.  Note that the different $B_{gc}$ cuts correspond to cuts in mass that are well described by a Gaussian with $\sigma_{\log_{10}}$ of 0.25 (red line).  The color scheme is the same as in Figure~\ref{Bgc scatter}.}
\label{Bgc cumscatter}
\end{figure}

\citet{rykoff08} report a best-fit relation between X-ray luminosity and their mass indicator ($N_{200}$), using the maxBCG clusters at the redshift from 0.1 to 0.3 \citep{koester07}.  $N_{200}$, their richness indicator, is given by the number of E/S0 ridgeline members falling within $R_{200}$ of the BCG and brighter than $0.4L_*$.  The best fit relation in their findings shows an intrinsic scatter $\sigma_{\ln L}=0.86 \pm 0.03$, which corresponds to $ \sigma_{\log_{10}L} \sim 0.37$.  That, in turn, corresponds to $\sigma_{\log_{10}M} \sim 0.25$, assuming the scaling relation between X-ray luminosity and mass is a power-law of the form log$_{10}L \sim\log_{10}M^{1.5}$ \citep{reiprich02}.  Figure~\ref{Bgc cumscatter} shows the cumulative distribution of $B_{gc}$ scatter in each $B_{gc}$ bin, and the cumulative Gaussian distribution with $\sigma$ of 0.25 (red dotted line).  Except for the highest $B_{gc}$ bin, the cumulative distribution is nearly Gaussian with $\sigma$ of 0.25.  These data support the use of Gaussian scatter in $\log_{10}(M)$ in the optical richness parameter for cosmological studies  \citep[e.g.,][]{gladders07}.  Interestingly, our measurement of the intrinsic scatter in the $B_{gc}$ mass relation is consistent with the intrinsic scatter estimated for $N_{200}$ that is extracted by leveraging the X-ray properties of galaxy clusters \citep{rykoff08}.  

\begin{figure}[ht]
\includegraphics[scale=0.45]{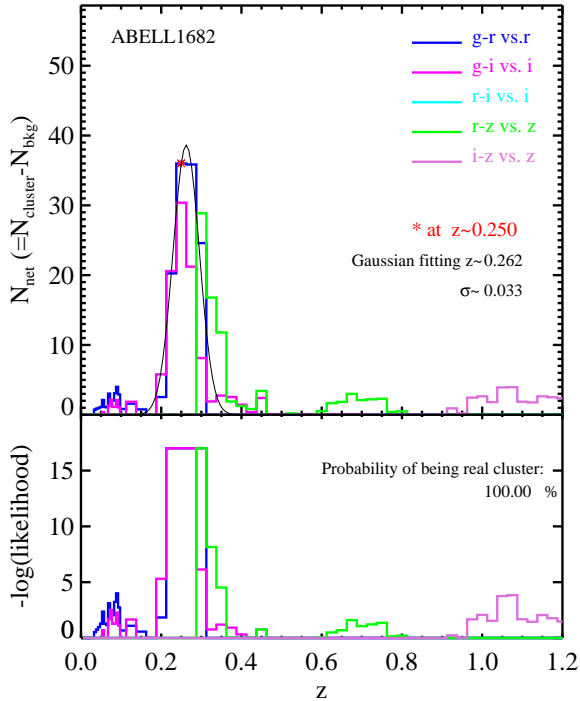}
\caption{The red sequence overdensity toward Abell 1682 ($z=0.226$) within 0.8Mpc from the X-ray center.  \emph{Top}: The number of galaxies within the area of 0.8Mpc radius from the center at each redshift and color which is corrected for an estimated number of background galaxies within the same aperture size at each redshift and color.  The solid black line is the fitted line assuming the overdensity distribution around the peak is Gaussian.  \emph{Bottom}: The likelihood for an overdensity to be a cluster signal compared to background noise is shown, which is then turned into the probability of being real cluster detection, assuming the background noise is Poissonian.}
\label{abell1682}
\end{figure}

\section{A Red-Sequence Redshift Estimator}
\label{photo-z}

In this section,  we present tests of a redshift estimator that has been used in studies of SZE and X-ray selected galaxy clusters \citep[][Suhada et al 2011, Williamson et al 2011]{staniszewski08,high10} and that forms the core of an optical cluster selection method being developed for joint optical and SZE cluster finding (Liu et al 2011).  In the case of an SZE or X-ray candidate, we examine galaxies within a region of the sky around the known center by choosing a physical radius of $r\sim0.8$~Mpc or an estimated virial radius $R_{200}$, depending on the application.  At each redshift we select only the galaxies with colors appropriate for the red-sequence, using the SSP model (BC model which is the same model that we use to paint red galaxies in the mock catalog described in \S\ref{galaxy colors}).  We apply a correction for the expected number of background galaxies with this color at each redshift and thereby end up with a measured overdensity of red--sequence galaxies as a function of redshift along the line of sight toward the X-ray or SZE cluster candidate.  All four available filters, $g$-, $r$-, $i$-, $z$-, are used to look for this red galaxy overdensity at all redshift range in order to avoid false peaks due to degeneracy between colors and redshift.  The degeneracies in colors of $g$-$r$, $r$-$i$ and $i$-$z$, has been known to generate false peaks at transitional redshift ranges (i.e., $\sim$0.35, $\sim$0.75 for z$<1$) where the 4000$\AA$ break moves out of one of the filters in colors.  By scanning through the data with five colors, $g$-$r$, $g$-$i$, $r$-$i$, $r$-$z$ and $i$-$z$, at certain redshift intervals, one can reinforce the real overdensity peaks to show up in any of these color combinations (sometimes in more than one colors), avoiding using $g$-$r$ at redshift around 0.3--0.35, for example.  

\begin{figure}[ht]
\includegraphics[scale=0.45]{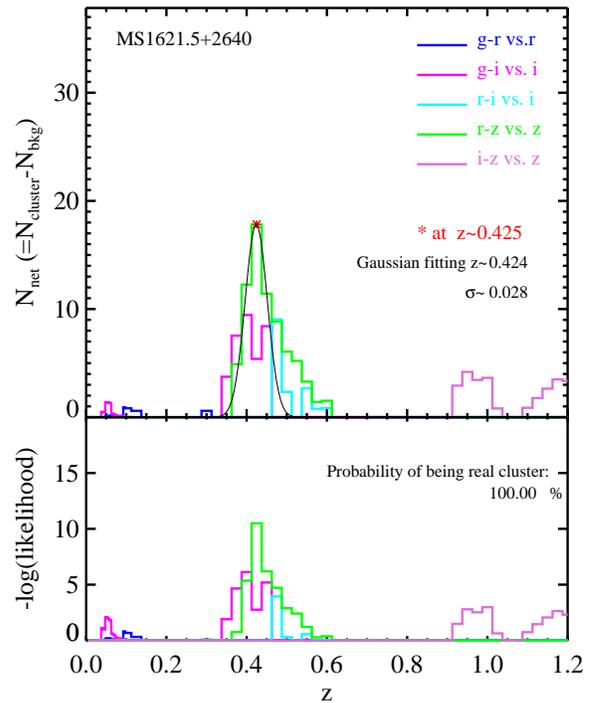}
\caption{The background corrected red-sequence galaxy overdensity for MS1621.5+2640 ($z=0.426$), same as in Figure~\ref{abell1682}.}
\label{ms1621}
\end{figure}

We have tested this approach using 51 X-ray and optically selected clusters that lie within the SDSS DR7 survey region.  For each cluster we estimate the cluster mass and virial radius using an X-ray temperature and the mass--temperature scaling relation from \citep{finoguenov01}.  Figure~\ref{abell1682} shows the results for Abell 1682 and Figure~\ref{ms1621} shows the results from the redshift estimation of MS1621.5+2640.  On the top panel of the histograms, the background corrected net number within the cluster's $r_{200}$ is plotted in each redshift bin.  Abell 1682 is a cluster with X-ray temperature of 7.24 keV at $z=0.226$ and MS1621.5+2640 has X-ray temperature of 7.6 keV at $z=0.426$.  Because the 4000~$\AA$ break in the old stellar populations of the cluster galaxies moves out of $g$-band at about $z=0.35$, the peak in the histogram for MS1621.5+2640 shows up in $r$-$z$ vs. $z$ at the appropriate redshift, while Abell 1682 shows a peak in $g$-$r$ vs. $r$ histogram at the appropriate redshift.  The bottom panel in each figure shows the likelihood of each detection where we assume Poisson noise.  The redshift bin with the maximum likelihood is present is chosen as the initial estimation of a cluster (indicated as red asterik on the top panel), and we refine the redshift estimation using a Gaussian function fit to the overdensity distribution around the initial peak.

\begin{figure}[ht]
\includegraphics[scale=0.45]{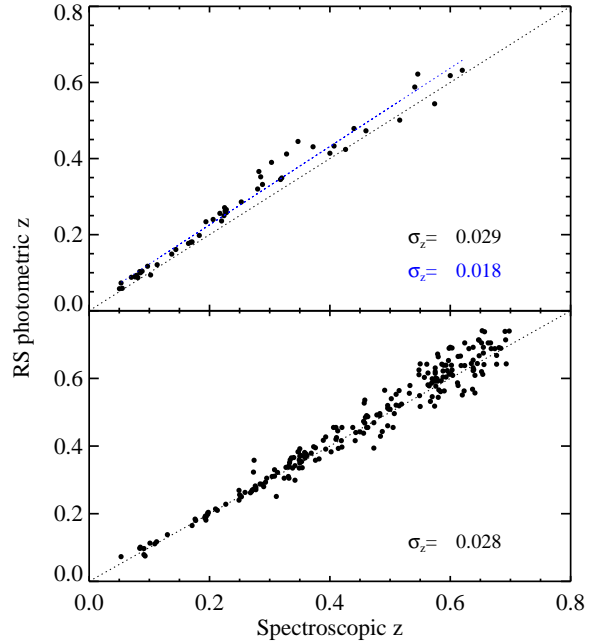}
\caption{Comparison between photometric and spectroscopic redshifts.  \emph{Top}: The comparison using an ensemble of galaxy clusters that overlap the SDSS DR7 data release.  The rms in blue shows the scatter from the best-fit line which is drawn in blue and includes a systematic overestimate shown in black in the photo-z's with the scatter of 0.029.  \emph{Bottom}: A comparable test is shown based on a galaxy catalog created by the galaxy simulator presented in this work at a similar survey depth of SDSS.}
\label{realzcmrz}
\end{figure}

In Figure~\ref{realzcmrz} on the top panel, we show a plot of the photometric redshift versus the spectroscopic redshift for the full ensemble of 51 clusters, while the bottom panel shows the same test using a simulated galaxy clusters described in this paper.  There is good overall agreement, with evidence that our estimates  systematically higher by with the root mean square (rms) scatter of the photo-z's around the true redshifts (black dotted line) of 2.9\% and an rms of 1.8\% once the bias is removed.  This systematic bias in redshift could be further reduced by additional tuning of the SSP models used to predict the red-sequence color and evolution, as supported by the test using the simulated galaxy cluster populations using the same SSP models for galaxy colors and the redshift estimator.

\section{Discussion}
\label{discussion}

We have presented an empirical method for constructing simulated catalogs that relies upon high resolution dark matter-only simulations and the observationally constrained properties of cluster and field galaxy populations.  This empirical approach is attractive because it offers the possibility to test, improve and characterize the final performance of optical cluster finders and other tools that are used on real galaxy catalogs.   This method can be further tuned as improved observational constraints on cluster and field galaxy populations become available.
 
In \S\ref{VTP} we have demonstrated the power of this approach by characterizing the selection of the VTP cluster finder.   We have used the mock catalog to measure the contamination as a function of redshift and the completeness as a function of mass and redshift.  The development version of this code performs reasonably well, with characteristic contamination of $\sim$40\% out to $z\sim0.55$, and completeness that increases with mass and reaches characteristic values of around 50\%.

An advantage of our method is the ease of modifying the galaxy populations by altering the population parameters such as the blue fraction mass dependence and redshift evolution, the intrinsic scatter in the red sequence, the evolution of the HON and the relative brightness of field and cluster galaxy populations.   We have used this property of the catalog simulator to extract the systematic uncertainties on the selection function using the full range of the catalog parameters that are consistent with current observational data.  Sensitivity of the completeness to variations in these parameters is at the $\sim$5\% level for blue fraction, HON and red sequence scatter changes, but it does approach $\sim$15\% at certain redshifts.  Uncertainties are only at the $\sim$1\% level due to relative brightness changes in the field and cluster LFs.  We also have demonstrated how one would combine effects of uncertainties in different parameters on estimating selection functions of a cluster finder, assuming the effects of parameters explored in this work are independent.  When they are all combined in quadrature, the uncertainty is at the level of $\le$15\%.  These uncertainties should be included in the cosmological analysis of optically selected cluster samples.  With additional work on the optical properties of cluster galaxy populations, especially in the high redshift regime, these uncertainties can be reduced to enable the full statistical power of the large optically selected cluster samples to be realized.

In addition, we used the ensemble of catalogs to test optical mass estimation (\S\ref{Bgc}) and redshift estimation (\S\ref{photo-z}).  Our analysis shows that the $B_{gc}$ optical mass estimator is correlated with cluster halo mass but with large scatter.  The scatter in mass at fixed $B_{gc}$ is approximately log normal and about 75\% ($\sigma_{\log_{10}(M)}\sim0.25$), which is markedly worse performance than X-ray and SZE mass estimators \citep{mohr99,ohara06,vikhlinin09,vanderlinde10,andersson10}.  The performance of the red sequence overdensity redshift estimator is better than 2\% once biases possibly associated with a mismatch between the evolution of observed and modeled red sequences are taken into account.

This project highlights the importance of empirical mock catalogs, not only for obtaining an accurate estimate of the selection function for a cluster finding algorithm, but also for characterizing the uncertainties in the selection function.  Moreover, a catalog generator is an essential tool during the development of tools for optical cluster finding, mass estimation and photometric redshift measurement.  The next step in this development stream is to extend to larger volume light cone outputs from other structure formation simulations and to extend the analyses of galaxy populations to higher redshift using deeper survey datasets such as those coming from the Dark Energy Survey. 
\acknowledgments
JS acknowledges the support of the DoE Grant DE-FG02-95ER40899.  JM acknowledges the support of the Excellence Cluster Universe in Garching.\\ \\

\bibliographystyle{apj}
\bibliography{mock_catalog}

\end{document}